\documentstyle{l-aa}
\input psfig.sty

\newcommand{\be}{\begin{equation} }
\newcommand{\ee}{\end{equation} }

\newcommand{\apj}{ApJ }
 
\newcommand{\apjs}{ApJ Supp.}
\newcommand{\mnras}{MNRAS }

\newcommand{\ltapprox}{\raisebox{-0.5ex}{$\,\stackrel{<}{\scriptstyle\sim}\,$}}
\newcommand{\gtapprox}{\raisebox{-0.5ex}{$\,\stackrel{>}{\scriptstyle\sim}\,$}}

\begin{document}

   \thesaurus{12         
              (11.03.1;  
               12.03.3;  
               12.04.1;  
               12.07.1)} 

\title{Two multiple-imaged $z = 4.05 $ galaxies in the cluster-lens 
Abell 2390 \thanks{Based on observations collected with the 
Canada-France-Hawaii Telescope (CFHT), 
the William Herschel Telescope (WHT) and the Hubble Space Telescope (HST).}} 

\author{ R. Pell\'o\inst{1}, J. P. Kneib\inst{1}, J. F. Le Borgne\inst{1},
J. B\'ezecourt\inst{1,4}, T.M. Ebbels\inst{2}, I. Tijera\inst{3},
G. Bruzual A.\inst{5,1}, J.M. Miralles\inst{1,6}, I. Smail\inst{7}, G. Soucail\inst{1},
\and T.J. Bridges\inst{2,8}}

\offprints{ R. Pell\'o
              roser@obs-mip.fr}

\institute{Observatoire Midi-Pyr\'en\'ees, UMR 5572,
14 Avenue E. Belin, F-31400 Toulouse, France
\and
Institute of Astronomy, Madingley Road, Cambridge CB3 0HA, UK
\and
Departament d'Astronomia i Meteorologia, Universitat de Barcelona,
Diagonal 648, 08028 Barcelona, Spain
\and
Kapteyn Instituut, PO Box 800,9700 AV Groningen, the Netherlands
\and
Centro de Investigaciones de Astronom\' \i a (CIDA), Apartado Postal 264, M\'erida, Venezuela
\and
Astronomical Institute, Tohoku University, Aramaki, Aoba-ku, Sendai 980-8578, Japan
\and
Department of Physics, University of Durham, South Road, Durham DH1 3LE, UK
\and
Anglo-Australian Observatory, PO Box 296, Epping NSW 1710 Australia
}

\date{Received, 1998, Accepted , 1998}

\maketitle
\markboth{R. Pell\'o et al.: Two $z \sim 4$ galaxies behind A2390}{ }

\begin{abstract}
We present the first results on the identification
and study of very distant field galaxies in the core of cluster-lenses,
using a selection criterium based on both lens modelling and photometric 
redshifts. We concentrate on two multiple-imaged sources at $z = 4.05$
in the cluster Abell 2390. 
The two objects presented in this paper,
namely H3 (cusp arc) and H5 (fold arc), were identified
through lens modelling as multiple images of high-redshift sources 
at $z \gtapprox3.5$ (Kneib et al. 1999). We confirm the excellent 
agreement between this identification and both their photometric redshifts 
and morphologies. Our CFHT/WHT program for a systematic redshift
survey of arcs in clusters has allowed to obtain a set of spectra on 
three different images at $z \sim 4 $: the brightest image of H3, 
whose redshift was already confirmed by Frye \& Broadhurst (1998),
and the two brightest images of H5. The later 
is then confirmed spectroscopically as a multiple image, giving a
strong support to the lens model. The main feature in each of these 
spectra is a strong emission line, identified as Ly $\alpha$, 
leading to $z = 4.05$ for both H3 and H5. The spectrophotometric 
properties of these galaxies are studied, in particular the degeneracy
in the parameter-space defined by the SFR type, age, metallicity and 
reddening. H3 and H5 are intrinsically bright and clumpy galaxies 
($M^{*}_B $ to $M^{*}_B $ -2 magnitudes), located $\sim 100 h^{-1}_{50}$ kpc apart 
on the source plane, with mean metallicities compatible with a fraction of 
solar or even solar values. These results seem to favour a hierarchical 
merging scenario, where we are seeing a relatively evolved phase in 
these two $z \sim 4 $ objects, with stars forming locally and
efficiently.

\end{abstract}

\section{Introduction}

The identification and study of very high-redshift galaxies
($z \ge 4$) is probably one of the most direct methods to constrain the 
scenarios of galaxy formation and evolution. The observation of such galaxies 
presents a major challenge complementing the
statistical studies of lower redshift galaxies (CFRS: Lilly et al. 1996;
Hawaii Deep Fields: Cowie et al. 1996;
HDF: Sawicki et al. 1997; Lowenthal et al. 1997; Steidel et al. 1996b). The key 
information to retrieve is the star formation history
in the universe and the evolution of the different morphological
types of galaxies. The main problems are the identification of high-z
galaxies and the construction of a sample as bias-free as possible. 
Steidel et al. (1992, 1993, 1996a, 1997) first used the Lyman dropout technique
to photometrically 
identify $3.0 \le z \le 3.5$ galaxies in empty fields, and they succeeded in 
confirming spectroscopically a large sample of them with the Keck telescope. A similar
technique has been used on the HDF to identify $2.2 \ltapprox z \ltapprox 3.5$
galaxies (Steidel et al. 1996b; Lowenthal et al. 1997) also 
confirmed with the Keck telescope. The number of massive star-forming galaxies
at high redshift can be used to constrain the cosmological models (see Baugh et
al. 1998). All these galaxies are in general very compact in their restframe UV, 
with the bulk of their 
star formation located on high surface-brightness regions. Their morphologies 
might give some hints on the physical processes involved in galaxy evolution. In
particular, they might provide useful constraints on the formation of the 
different systems (spheroids, bulges, disks), and its close relation with metal
enrichment timescales (see Trager et al. 1997 for a detailed discussion). At
$z \sim $ 3 to 4, the corresponding lookback time is 13.7 to 14.4 Gyr, 15 Gyr
being the present age of galaxies ($H_0 = 50$ km s$^{-1}$ Mpc$^{-1}$, $q_0 = 0.1$ 
and $\Lambda = 0$). Thus, such high redshift systems provide a direct measure 
of the early star-formation processes. 

The photometric technique based on the identification of Lyman dropouts in the U-band 
has shown to be a useful tool to select $2.2 \ltapprox z \ltapprox 3.5$ galaxies, and it
has been successfully extended to the B-band to locate $3.5 \ltapprox z \ltapprox 
4.5$ candidates (Steidel et al. 1998). Nevertheless,
two major biases will appear in all such selection techniques: one towards intrinsically 
luminous galaxies because the sample is limited in apparent magnitude, and another towards 
galaxies with strong star-formation activity, because the rest-frame UV will be seen 
at the visible wavelengths. When high-redshift galaxies are selected through 
photometric redshift techniques based on a large wavelength domain,
including near-IR, and close to critical-lines of cluster-lenses,
the resulting sample is less sensitive to these biases. This paper presents 
the first $z \sim 4$ results on this 
original method to build-up and study an independent sample of $z \gtapprox 
2 $ galaxies, by combining photometric redshifts (visible and near-IR) and 
lens modelling. One of the first examples of such a technique on lensed
galaxies is the $z=2.51$ star-forming object in A2218 (Ebbels et al. 1996), 
and a number of recent discoveries of $z \gtapprox 2.5$ 
lensed-galaxies in clusters, sometimes serendipituous, strongly encourages 
this approach (Yee et al. 1996a; Trager et al. 1997; Franx et al. 1997; 
Soifer et al. 1998; Seitz et al. 1998;
Frye \& Broadhurst 1998; Bunker et al. 1998). 

The two $z \sim 4$ lensed galaxies discussed in this paper were identified in the
cluster Abell 2390 ($z=0.231$), a gravitational lens which has been extensively 
studied. The first detailed photometric and spectroscopic 
surveys on this target were performed by Le Borgne et al. (1991), and recently 
enlarged by the CNOC group (Yee et al. 1996b; Abraham et al. 1996;
Carlberg et al. 1996). This cluster, which is known as a strong X-ray emitter
(Ulmer et al. 1986; Pierre et al. 1996), has an elongated shape
($\epsilon = (a-b)/(a+b) = 0.55 $, based on the
luminosity map of the photometrically selected cluster-galaxies, Pell\'o et al. 1991)
and a high velocity-dispersion ($1090 $km/s, Abraham et al. 1996). 
A near-IR photometric survey is also available (Miralles et al. 1999, in
preparation). We will focus here on two multiple images at $z \sim 4 $,
namely H3 and H5, according to the nomenclature by Kneib et al. (1999, 
see also Figure 1). The redshift of the former had already been confirmed 
by Frye \& Broadhurst (1998) at the Keck telescope. 

A plan of the paper follows.
First, we give a summary of the spectroscopic and photometric data in Section 2.
Section 3 is devoted to the study of the main properties 
of H3 and H5, namely the spectroscopic and photometric redshifts, the 
lensing properties and the morphology of these objects.
The stellar population of these objects is characterized in Section 4. 
In Section 5 these results are discussed, with a special attention to 
the implications for galaxy 
formation and evolution. Throughout this paper, we assume $H_0 = 
50$ km s$^{-1}$ Mpc$^{-1}$ and $\Lambda = 0$.

\section{Observational data}

\subsection{Photometry}

The photometric data were obtained during a number of runs since 1988, at
the 3.6m Canada-France-Hawaii Telescope, the 3.5m telescope of Calar Alto 
(CAHA, Spain) and the 2.5m Isaac Newton
Telescope (Spain). We have selected the images obtained under photometric
conditions. The total number of filter bands considered is 10
when we add the two F555W ($V_{W}$) and F814W ($I_{W}$) WFPC2 images
from our HST program (PI. B. Fort), although the two B bands have very similar 
central wavelengths (but different widths). Table 1 provides with the central
wavelengths and widths of the different filters. In the case of the 2 pairs
(R and r, I and $I_{W}$), the difference between the central wavelengths is of
the order of $\sim 300 \AA$, and the filter widths are different. We use all
these data as complementary information in the following discussion, and we
consider different independent sets of filters to constrain the properties 
of the stellar population (see Sect. 4).
All magnitudes are given in the Vega system, and we use the 
SED of Vega and the detailed filter and detector responses to set the zero 
points when converting magnitudes into fluxes.
As most of these data have been 
published elsewhere, we give only a summary of the main characteristics 
of the images in Table 1, including the detection levels and references. 
The $2 \sigma$ magnitude is defined here as the magnitude of an object 
with 4 connected pixels $2\sigma$ above the sky level.

\begin{table}
\caption[]{Characteristics of the images and detection levels.
$2 \sigma$  magnitudes correspond to objects with 4 connected pixels,
each 2 $\sigma$ above the sky level. Observations were carried out at 
the CFHT (1), the INT (2), HST (3) and CAHA (4). }
\begin{flushleft}
\begin{tabular}{lllrrrrrll}
\hline\noalign{\smallskip}
  & $t_{exp}$ & $\sigma$ & pix
& $\lambda_{eff}$ & $\Delta\lambda$ & $ m $ & Ref. \\
  &(ksec)&(\arcsec)&(\arcsec)&(nm)&(nm)&
$2 \sigma$ &   \\
\noalign{\smallskip}
\hline\noalign{\smallskip}
B      & 5.4& 0.88 & 0.21 & 436 &  70 & 26.7 & a, 1 \\
$B_{J}$& 2.7& 0.68 & 0.21 & 437 & 107 & 26.8 &  1 \\
g      & 2.1& 1.0  & 0.74 & 486 &  40 & 24.7 & b, 2 \\
$V_{W}$& 8.4& 0.13 & 0.10 & 545 & 105 & 28.5 & c, 3 \\
R      & 2.7& 0.74 & 0.21 & 641 & 141 & 25.2 &  a,1 \\
r      & 5.4& 0.80 & 0.25 & 669 &  64 & 25.8 &  4 \\
$I_{W}$&10.5& 0.13 & 0.10 & 799 & 137 & 27.7 & c, 3 \\
I      & 4.2& 1.0  & 0.42 & 832 & 113 & 23.8 & 1 \\
J      & 5.3& 1.1  & 0.50 & 1237& 147 & 23.0 & d, 1 \\
K'     & 4.1& 1.1  & 0.50 & 2103& 272 & 21.4 & d, 1 \\
\noalign{\smallskip}
\hline
\end{tabular}
\begin{tabular}{ll}
a) Pell\'o et al. 1991 \\
b) Le Borgne et al. 1991 \\
c) Kneib et al. 1999 \\
d) Miralles et al. 1999, in preparation. \\
\end{tabular}
\end{flushleft}
\end{table}

\subsection{Spectroscopy}

The spectroscopic data set used in this paper comes from three 
different runs: one at the CFHT 
and two at the 4.2m William Herschel Telescope (WHT). H3 was one of the 
targets of the spectroscopic survey performed by B\'ezecourt and 
Soucail (1997) at CFHT in August 1995. The spectrograph used was MOS/SIS 
(Le F\`evre et al., 1994), with the V150 grism, providing a low dispersion
of $7.3 $ \AA /pixel. 
H5a and H5b were observed during two separate runs at the WHT, 
in September and June 1996 respectively, with the LDSS-2 multiobject 
spectrograph (Allington-Smith et al. 1992). The grism used 
was the medium blue, with a dispersion of $5.3 $ \AA /pixel.
The total exposure times were 6.7 ksec 
($2 \times 2500$ sec $+$ 1700 sec) on H5b and 9.9 ksec 
($2 \times 3600$ sec $+$ 2700 sec) on H5a.
The reduction was performed using standard 
IRAF procedures as well as our own software packages.

\section{Photometric and spectroscopic study of H3 and H5}

\subsection{Lensing properties, spectral energy distribution and morphology of the sources}

Hereafter we use the new lens model of A2390 by Kneib et al. (1999), which is
a refined version of the earlier model presented in Pierre et al. (1996), 
based on lensing and X-ray data, taking into account the new constraints given by the HST images.
According to this lens modelling, H3 and H5
are multiple images of two high-redshift sources at $z \gtapprox 3.5$.
This redshift estimate has been confirmed through photometric redshift
techniques (filters from B to K) and spectroscopy.
Figure 1 displays a zoom on the different components of H3 and
H5, as well as their location in the cluster. H3a-b-c 
is an impressive cusp arc showing 
several bright knots which are identifiable in each different image.
H5 is a fold arc with the two radial components showing similar
morphologies. A third faint image is predicted for this source,
but unfortunately it lies on the edge of the Planetary Camera, where 
the exposures have a poor
S/N. Table 2 summarizes the photometry of all these 
components except H3c, which is too close to the bright 
cluster galaxy. The different colors have been obtained through 
aperture magnitudes computed within the same region in all the
filters, correcting for sampling and seeing effects. The averaged colors 
of a bright E/S0 cluster galaxy are also given for comparison. 
The surface brightnesses and colors are
compatible with the multiple-image hypothesis within the 
photometric errors ($\Delta m \sim 0.1 $ typically, 
and $\Delta m \sim 0.3 $ in the near-IR). 
The faint images H5a and H3b are surrounded by bright objects
(see Fig. 1), and they are hardly detected in the near-IR and g
where the sampling ($\ge$ 0\farcs5 per pixel) and/or the seeing
conditions are poor. Moreover
H5a and H5b are hardly detected in B. In these particular cases, 
photometric errors are at least $\Delta m \sim 0.5 $, and these magnitudes
are indicated by ":" in Table 2. It is worth 
noting that the identification of multiple images based on the 
similarity of the spectral energy distributions (SED) 
is more easily obtained with extended wavelength coverage.

\begin{figure*}
\psfig{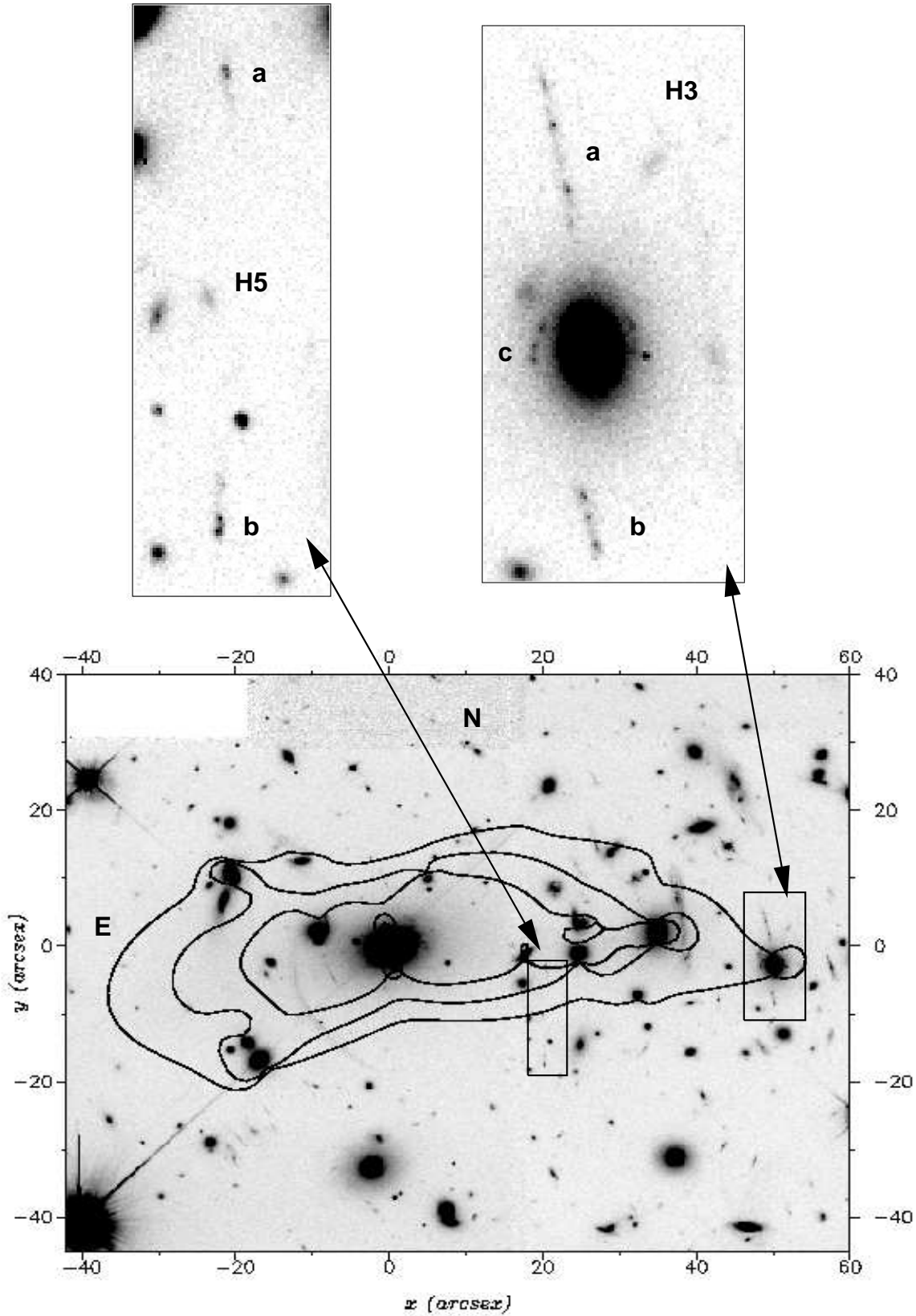}
\caption{Image of Abell 2390 in logarithmic scale, 
obtained by coadding the final HST images in $V_{W}$
and $I_{W}$. A zoom on H3 and H5 is also shown, with the 
identification of the different components. The critical lines
at $z =$ 1, 2.5 and 4 (from the inner to the outer part, respectively)
are also displayed, according to the model by Kneib et al. (1999).}
\label{}
\end{figure*}

\begin{table}
\caption[]{Photometry of the different H3 and H5 components, compared
to the mean values for E/S0 cluster galaxies (see text).}
\begin{flushleft}
\begin{tabular}{rrrrrrr}
\hline\noalign{\smallskip}
Filter & H3a & H3b & $\Delta m_{ab}$ & H5a & H5b & $\Delta m_{ab}$ \\
\noalign{\smallskip}
\hline\noalign{\smallskip}
B      &  -   &   -  &  -  & $>$27.1& $>$25.6& -  \\
$B_{J}$    & $>$25.5 &  -   &  -  & $>$29.2& $>$27.0& -  \\
g      & 24.4 & 25.3:& 0.9:&$>$25  & 25.2 &$>$0.2 \\
$V_{W}$& 23.8 & 24.1 & 0.3 & 25.2 & 24.4 & 0.8 \\
R      & 22.8 & 23.0 & 0.2 & 23.5 & 23.1 & 0.4 \\
r      & 22.9 & 23.4 & 0.5 & 23.9 & 23.1 & 0.8 \\
$I_{W}$& 22.1 & 22.5 & 0.4 & 23.7 & 23.0 & 0.7 \\
I      & 22.1 & 22.3 & 0.3 & 23.3 & 22.5 & 0.8 \\
J      & 21.0 & 21.6 & 0.6 & 23.0:& 22.3 & 0.7:\\
K'     & 19.7 & 20.2 & 0.5 & 21.9 & 21.1 & 0.8 \\
\noalign{\smallskip}
\hline
\hline
\end{tabular}
\begin{tabular}{rrrcrr}
        & g-r & $V_{W}$-R & $V_{W} - I_{W} $ & I - J & J - K \\
H3a     & 1.5 &  1.5      &  2.1             & 1.0   & 1.3 \\
H3b     & 1.9:&  1.8      &  2.4             & 0.7   & 1.4 \\
H5a     & $>$1.1& 1.7     &  1.5             & 0.3   & 1.1 \\
H5b     & 2.1 & 1.3       &  1.4             & 0.2   & 1.2  \\
cluster & 1.3 & 0.8       &  1.7             & 1.3   & 1.1  \\
\noalign{\smallskip}
\hline
\end{tabular}
\end{flushleft}
\end{table}

The first spectrum of H3 obtained by B\'ezecourt and Soucail (1997) shows a single
emission line which was assigned to [O II]$3727 $ \AA, giving $z=0.647$,
a value which is in fact incompatible with both the lens modelling and the 
photometric redshift. On the contrary, when it is correctly assigned to the 
rest-frame Ly$\alpha$, it gives $z = 4.05$. Subsequent observations of H3 by 
Frye \& Broadhurst (1998) at the Keck Telescope obtained a redshift of 
$z = 4.04 $ through a higher quality spectrum. This time, the redshift is based on 
several absorption lines, the Ly$\alpha$ centroid being slightly shifted redwards with
respect to the absorption system, in good agreement with the position of the 
line found by B\'ezecourt and Soucail (1997).

In the case of H5a and H5b, both spectra
show a strong emission line (Fig. 2). When this line is identified as Ly$\alpha$, 
the corresponding redshifts are $z = 4.049\pm 0.003$ and $z = 4.052\pm 0.002$ for 
the a and b components respectively. Thus, this result confirms that H5a and H5b
are two images of the same $z = 4.05$ source, in perfect agreement with the lens 
model. When taking the Ly$\alpha$ centroid to compute the redshift, the
two multiple images H3 and H5 are at the same redshift within the errors.
Hereafter we take the Ly$\alpha$-based redshift of $z = 4.05$ for both sources,
because we have no estimate based on absorption lines for H5. The measured 
equivalent width of Ly$\alpha$ is relatively high for H5,
$W_{\lambda} = 273 $ \AA\ for the brightest image, corresponding to a rest frame value 
of $54 $ \AA. We obtain a rest frame upper limit of $\sim 75 $ \AA\ for the faintest image,
where the continuum is hardly detected. These values are similar to the ones 
observed by Hu et al. (1998) in their sample of emission-line galaxies at
$z \sim 3$ to 6. The equivalent width of H3 is smaller, the upper limit obtained from the 
CFHT spectrum being $\sim 33 $ \AA. 

The gravitational amplification computed for the brightest images H3a and 
H5b is $2.3 \pm 0.3$ magnitudes in both cases, and the predicted difference 
between the two images are $\Delta m_{ab} $ = 0.4 and 0.8 magnitudes for
H3 and H5 respectively. The corresponding measured differences (Table 2) are
also consistent with this model, $\Delta m_{ab} (H3)= 0.4 \pm 0.13 $ and 
$\Delta m_{ab} (H5)= 0.7 \pm 0.16 $. These amplification factors
are surface-averaged values computed with the Kneib et al. (1999) lens model.

We have applied a lens inversion procedure to restore the morphology of these
sources on the source plane at $z = 4.05$. This method is close to the
LensClean algorithm (Kochanek \& Narayan 1992; Kneib et al. 1994).
Figure 3 displays the resulting morphology and location of the
two objects on their source plane. The true separation between H3 and H5 
in the source plane would be $16''$, 
corresponding to a linear separation of $150 (252) h_{50}^{-1}$ kpc with 
$q_0= 0.5 (0.1)$. Note that these sources are not resolved in 
their width, the lens inversion being limited by the resolution of the
composite HST images ($0.13''$ in $V_{W}$ and $I_{W}$).
Both sources are extremely clumpy. H5 consists of an alignment of 
two compact and bright blobs, less than $0.1''$ apart ($0.9 (1.6) h_{50}^{-1}$ kpc), and a faint 
extended component of $\sim 0.3''$ ($2.8 (4.7) h_{50}^{-1}$ kpc), the total length being 
$\sim 0.4''$ ($3.7 (6.3) h_{50}^{-1}$ kpc). H3 is more 
elongated than H5, and it displays four small and bright subclumps.
Each one of these subclumps has less than $0.1''$ of diameter, and the total length
of the structure is about $\sim 0.8''$ ($7.5 (12.6) h_{50}^{-1}$ kpc). The orientation of the
two sources is similar (see Fig. 3). Compared to the morphologies and sizes 
of the $2.5 \ltapprox z \ltapprox 3.5$ sample by Steidel et al. (1996a), typically
$\sim 0.5''$ - $1''$ for the resolved cores ($\sim 3.6$ to $7.3h_{50}^{-1}$ kpc 
with $q_0=0.5$; see also Giavalisco et al. 1996),
the bright subclumps of H3 and H5 are more compact, but the total length 
of the emitting region is of the same order. From the morphological point of view, 
H3 and H5 are similar to the $3.3 \ltapprox z \ltapprox 4.0$ lensed sources found 
behind Cl0939+4713 (Trager et al. 1997).

\begin{figure}
\psfig{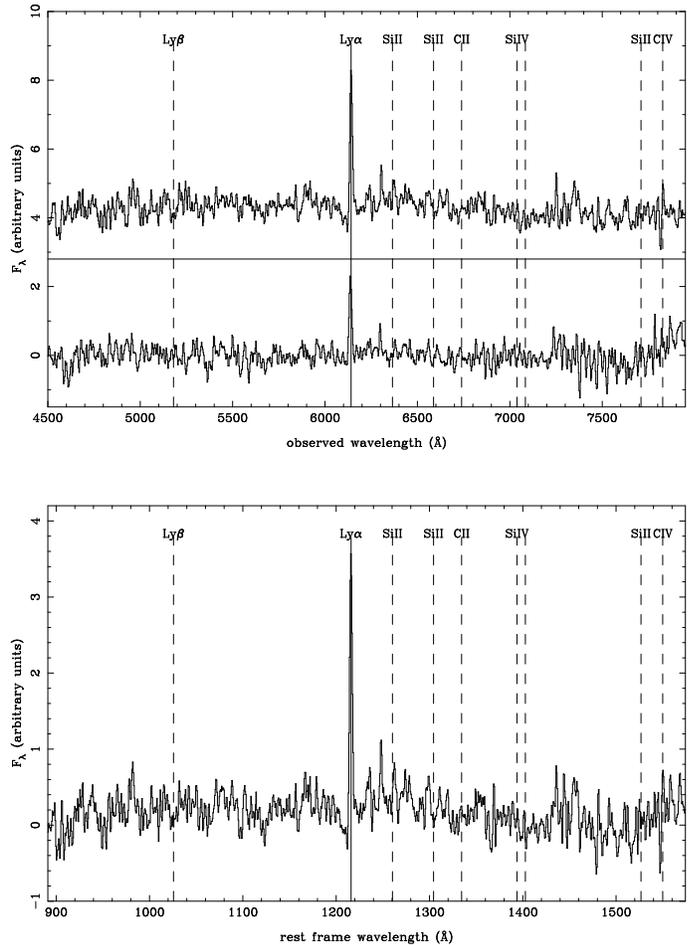}
\caption{Top: Mean spectra of the b (top) and a (bottom) components of H5, each of
them showing a strong emission line. Bottom: Averaged spectrum of the two H5 components,
showing the emission line identified as Ly$\alpha$ at $z = 4.05$. No other features are
clearly identified on the continuum.
}
\label{}
\end{figure}

\begin{figure*}
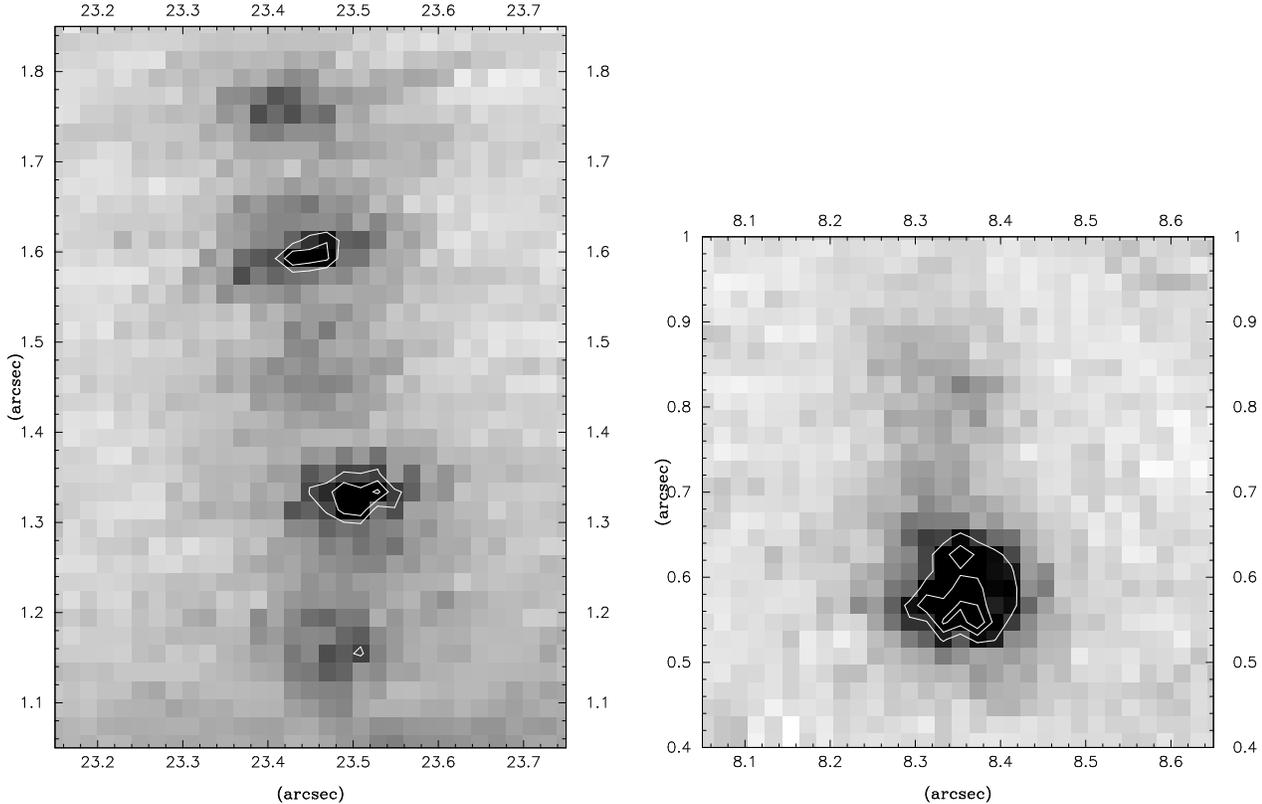

\hbox{
\psfig{file=8256.f3a,width=8.5cm}
\psfig{file=8256.f3b,width=8.5cm}
}
\caption{Restored $I_{W}$ image of the two objects H3 (left) and H5 (right)
on the source plane at $z = 4.05$, as obtained from the lens-inversion
procedure. Both sources are clumpy, elongated and 
exhibit the same orientation. The distance between H3 and H5 is $16''$.
At $z = 4.05$, $1''$ corresponds to a linear separation of $9.4 (15.7) 
h_{50}^{-1}$ kpc with $q_0= 0.5 (0.1)$. Isocontour plots display the 
center and profile of the different clumps, according to 
a linear scale which is identical for the two objects.}
\label{}
\end{figure*}

\subsection{The photometric redshift approach}

H3 and H5 are the first two high-redshift objects spectroscopically confirmed in 
our sample of photometrically selected candidates at $z \ge 2 $ in this
cluster. A photometric redshift method has been used to identify such candidates
in the cluster core, close to the critical lines. Photometric redshifts were
derived according to the standard minimization method described by Miralles, Pell\'o \& 
Roukema (1999) and Miralles (1998). The observed SED of each galaxy, as obtained
from its multicolor photometry, is compared to a set of template spectra.
The new Bruzual \& Charlot evolutionary code (GISSEL98, Bruzual 
\& Charlot, 1993, 1998) was used to build 5 different synthetic star-formation 
histories, each with solar metallicity ($Z_{\odot}$): a burst of 0.1 Gyr, a constant 
star-formation rate, and 3 $\mu$ models (exponential-decaying SFR) with characteristic time-decays 
matching the present-day sequence of colors for E, Sa and Sc galaxies. The template database 
includes 255 
synthetic spectra. The intergalactic absorption in the Lyman forest is modelled
using the average flux decrements $D_A$ and $D_B$, according to the original definition by
Oke \& Korycansky (1982). $D_A$ and $D_B$ correspond respectively to the continuum depression 
between Ly$\alpha$ and Ly$\beta$, and between Ly$\beta$ and the emission Lyman limit. 
The prescriptions for the redshift distribution of
$D_A$ and $D_B$ are taken from Giallongo \& Cristiani (1990), and they are in
good agreement with those given by Madau (1995) in the common redshift
domain $2.5 \le z \le 5$.
When applied to our data, the photometric redshift method identifies $\sim 30$ sources 
at $z \ge 2 $ in this field, most of them too faint to be confirmed spectroscopically using 4m 
telescopes. The high gravitational amplification of H3 and H5, and their strong
Ly$\alpha$ emission makes it possible in these
two particular cases. The results obtained on the whole field of A2390 and
two other cluster-lenses
will be discussed in more details elsewhere (see a preliminary version in Pell\'o et 
al., 1998). 

In the case of H3 and H5, we have examined the sensitivity of our photometric redshift 
estimate as a function of the relevant parameters, namely SFR, age and metallicity of the 
stellar population. This exercise is especially important for
H5, because the spectroscopic redshift is based on a single line. Figure 
4 presents a likelihood map for this object, showing a good agreement between the 
spectroscopic ($z=4.05$) and the photometric redshift ($z=3.90^{-0.38}_{+0.21}$,
where the error bar corresponds to a $1 \sigma$ level). 
This map was obtained using the set of 10 different 
SFRs presented in the next section, representing templates with a range of metallicities.
Each point on the redshift-age map corresponds to the best fit of the SED obtained 
across the SFR-metallicity space. The dark regions in Figure 4 (above $99 \%$ confidence level)
result from the overlap of the 5 metallicities considered (age vs. metallicity degeneracy).
The redshift region around $z \sim 4$ appears as the 
most likely solution for this object. The map obtained when using the 
solar-metallicity set above mentioned is qualitatively the same. In the case of H3,
we obtain a photometric redshift of $z=3.80^{-0.56}_{+0.60}$ using the same procedure, 
thus in good agreement with the spectroscopic value, but with a larger error bar.

\begin{figure}
\psfig{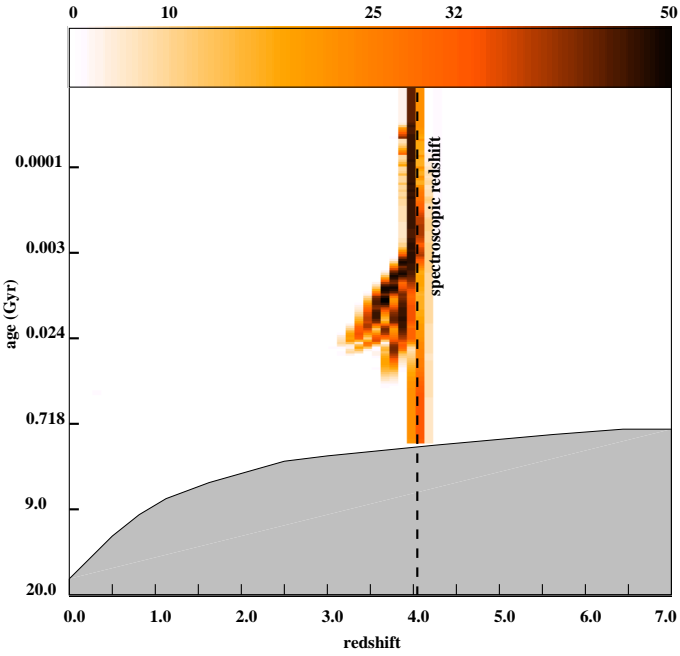}
\caption{Photometric redshift likelihood-map of H5 showing the excellent agreement
with the spectroscopic redshift of this object, which is contained within the 
region at 68\% confidence level. The scaling displayed at the top represents the
likelihood value associated to the $\chi^2$.
The shaded area enclose the $3 \sigma$ contour 
(confidence level of $99 \%$, or a likelihood value of $1 \%$). The shaded region on 
the lower part of the map is 
excluded because of age-limit considerations for the stellar population (stars cannot
be older than the age of the universe, with  $H_0$ = 50 km s$^{-1}$ Mpc$^{-1}$ and 
$q_0=0$).}
\label{}
\end{figure}

\section{Constraining the stellar population in H3 and H5}

We use the SEDs of H3 and H5, determined from broad-band photometry,
to infer the relevant parameters for the stellar population, in particular 
the total SFR. The SEDs of these objects can be 
fitted by different synthetic stellar populations, and there is a degeneracy 
to consider in the SFR-age-metallicity-reddening space. Figure 5 displays a comparison
between the SEDs of H3 and H5. The wavelength interval sampled by the broad-band filters in
the restframe of H3 and H5 goes from the Lyman 912~\AA \ break to $\sim 4400~$ \AA.
The region at 1500~\AA \ is sampled by the $I_{W}$ band, and the bands from R to J give
the slope of the UV continuum, whereas the K' band is sensitive to
the Balmer and 4000~\AA \ breaks (roughly the restframe B). Thus, we have in principle 
a good database to roughly constrain the stellar population in these objects. 
When the IMF and the upper mass limit for star-formation are fixed, 
the allowed parameter space can be roughly constrained. 
The presence of Ly$\alpha$ in emission points towards a star-forming system.
Again, we
used the GISSEL98 code for this exercise, taking into account that the two galaxies are 
necessarily dominated by massive OB stars at the wavelengths seen in the visible bands. 
Two kinds of SFRs were considered: a single stellar population (instantaneous burst), 
and a continuous star-forming system, both with the Salpeter (1955) IMF,
with upper and lower mass-cutoff of $0.1M_{\odot} \le m \le 125 M_{\odot}$,
and an extinction law of SMC type given by Pr\'evot et al. (1984). When computing quantities 
related to the stellar mass involved in one of these regions, we take the above 
mass limits for star-formation, but only $\sim 1/3$ of the stellar mass corresponds to
stars with $m \gtapprox 1M_{\odot}$ with this particular IMF.

\begin{figure}
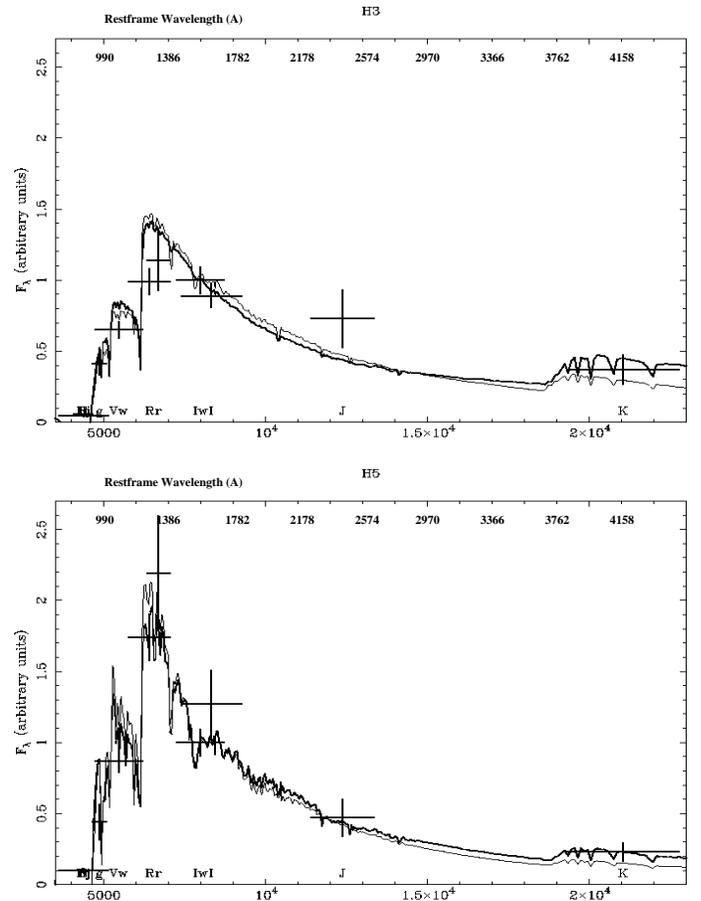

\vbox{
\psfig{file=8256.f5a,angle=270,width=9.cm}
\psfig{file=8256.f5b,angle=270,width=9.cm}
}
\caption{Spectral energy distribution of H3 (top) and H5 (bottom),
as determined from broad-band photometry. Fluxes are arbitrarily normalized 
to the mean flux in the $I_{W}$ filter, and we use the same flux-scale for 
both sources. The best fit constant star-formation and burst models are 
displayed by thick and thin lines respectively.
For H3, the best constant SFR has
0.02 $Z_{\odot}$ metallicity, $A_v=0.35$ and
age 2.4 Gyr ($\chi^2=0.85$), and a burst model has
0.2 $Z_{\odot}$ metallicity, $A_v=0.60$ and age 0.004 Gyr ($\chi^2=0.91$).
For H5, the equivalent values are 
$2.5 Z_{\odot}$, age 0.51 Gyr ($\chi^2= 0.83$) for the  constant SFR, and
$Z_{\odot}$, age 0.012 Gyr ($\chi^2= 0.99$) for the burst, all of them 
with $A_v=0.0$.
See text for more details.}
\label{}
\end{figure}

Figures 6a and 6b show the likelihood maps corresponding to H3 and H5 respectively,
at z=4.05.
For each SFR model, we have computed a reduced $\chi^2$ for 220 different ages of the
stellar population, 61 extinction values (ranging between $A_v=0.0$ and 3.0
magnitudes), and 5 different metallicities ($0.02 Z_{\odot}$, $0.2 Z_{\odot}$,
$Z_{\odot} = 0.02$, $2.5 Z_{\odot}$ and $5Z_{\odot}$). 
The most probable regions in this 
parameter-space are displayed in dark (permitted regions). The scaling directly 
corresponds to the confidence level as derived from the $\chi^2$ value. The shaded 
regions enclose the $3 \sigma$ contours (confidence level of $99 \%$). The likelihood maps 
were computed using two independent sets of filters: g,$V_{w}$,R,$I_{w}$,J,K and 
g,$V_{w}$,r,$I_{w}$,J,K, and the averaged intergalactic absorption in the Lyman forest 
(see Sect. 3.2). In all the cases we use $I_{w}$ instead of I because the seeing and
sampling are better in this filter. The results obtained with each independent set of 
filters are qualitatively the same, and the two $1 \sigma$ regions in the parameter space 
coincide. Each point on the likelihood maps presented in Figures 6a and 6b corresponds to the 
most restrictive value obtained from the two independent sets (i.e. the lowest 
likelihood value).
We also discuss in each particular case the results 
obtained when using only the spectral region redwards of Ly$\alpha$ (hereafter 
redSED), a region which is not affected by uncertainties on the intergalactic absorption. 
In order to retain good resolution at short time scales, a logarithmic scale is used to display 
the age of the stellar population.

According to Figures 6a and 6b, the stellar population in H5 is better constrained than 
in H3 in terms of $A_v$. The reason for this is probably the relatively redder spectrum of H3 compared to H5. 
The weight of the old stellar population is more
important in H3 than in H5 and, as a consequence, we expect an enlarged set of permitted
regions on the reddening-age plane. In a short-burst model without reddening,
H3 should be older than H5. Figures 6a and 6b also display the
degeneracy in metallicity versus reddening for the two objects. 
H3 and H5 do not seem to be highly reddened: the maximum restframe $A_v$ ever attained
at a $3 \sigma$ level is $\sim1.1$ magnitudes for H3, and it remains below $\sim0.8$ magnitudes
at $1 \sigma$. This is not surprising given the redshift of these objects and the 
photometric selection. The age of the stellar population in the constant 
star-formation models is unconstrained at $3 \sigma$ level in all the cases.
Table 3 summarizes the permitted domains in the parameter 
space for H3 and H5.

\begin{figure*}
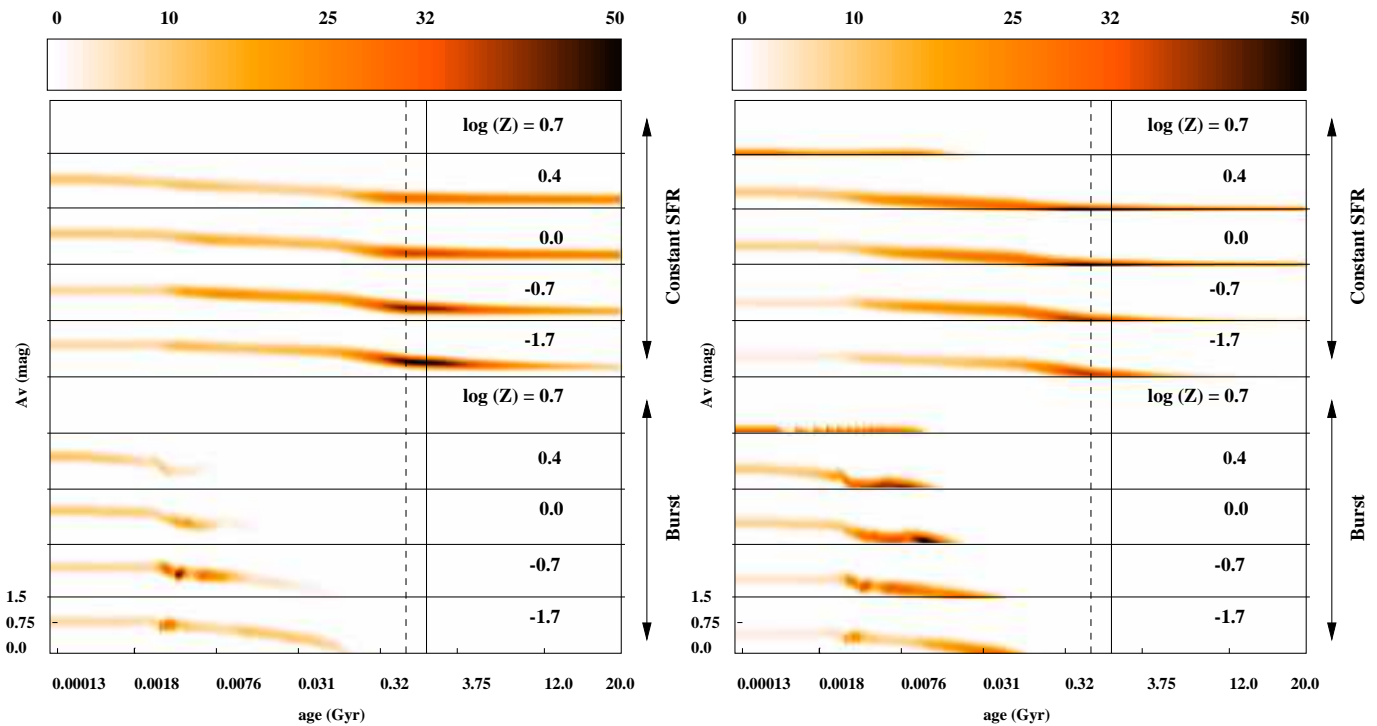

\hbox{
\psfig{file=8256.f6a,width=9.0cm}
\psfig{file=8256.f6b,width=9.0cm}
}
\caption{Likelihood map of H3 (left) and H5 (right) showing in dark the most probable regions and the 
degeneracy in the parameter space defined by SFR type, age, metallicity and reddening. 
Dotted and solid lines give the age limit corresponding to z=4.05, with $q_0=0.5$ and 0.1 respectively. 
Metallicities are given in units of log(Z)=log(Z/$Z_{\odot})$, and the solar metallicity is
$Z_{\odot} = 0.02$. The shaded regions enclose the $3 \sigma$ contours (confidence level 
of $99 \%$, or a likelihood value of $1 \%$, according the scaling displayed at the top).
A logarithmic scale is used for ages in order to retain good resolution at short time
scales. }
\label{}
\end{figure*}

\begin{table*}
\caption[]{Permitted domains in the parameter space of H3 and H5. The values
given are rough limits, the correlation between the different parameters is
displayed in figures 6 and 7.}
\begin{flushleft}
\begin{tabular}{llrrrrrr}
\hline\noalign{\smallskip}
 & Model & $1 \sigma$ & & & $3 \sigma$ & & \\
\noalign{\smallskip}
\hline\noalign{\smallskip}
\hline\noalign{\smallskip}
 &   & Age (Gyr) & $A_v$ (mags) & Metallicity ($Z_{\odot})$& Age (Gyr) & $A_v$ (mags) & 
Metallicity ($Z_{\odot}$)\\
\noalign{\smallskip}
\hline\noalign{\smallskip}
H3  & Burst        & 0.003 - 0.007 & 0.5 - 0.8 & 0.02 - 0.2 & 0.0 - 0.10 & 0.0 - 1.1 & 0.02 - 2.5 \\
H3  & Constant SFR & 0.18  - 1.17(2.16) & 0.2 - 0.6 & 0.02 - 1.0 & 0.0 - 1.17(2.16)& 0.1 - 1.1 & 0.02 - 2.5 \\
H5  & Burst        & 0.003 - 0.016 & 0.0 - 0.25 & 0.2 - 5.0 & 0.0 - 0.07 & 0.0 - 0.7 & all  \\
H5  & Constant SFR & 0.0001- 1.17(2.16) & 0.0 - 0.15 & all & 0.0 - 1.17(2.16)& 0.0 - 0.6 &  all  \\
\noalign{\smallskip}
\hline
\hline
\end{tabular}
\begin{tabular}{l}
\end{tabular}
\end{flushleft}
\end{table*}

H3 is well fitted by both the burst and the constant star-formation models. 
Within the $99\%$ confidence level, the $5 Z_{\odot}$ metallicity is excluded,
whatever the star-formation model, the reddening or the age of the stellar population.
All the solutions are compatible at $1 \sigma$ with a moderate $A_v \sim $ 0.4 to 0.6
magnitudes. The best fit at $1 \sigma$ with a burst model is obtained for metallicities
below solar: 0.2 $Z_{\odot}$ metallicity, $A_v=0.60$ and age 0.004 Gyr ($\chi^2=0.91$)
At $3 \sigma$, the maximum age for the burst model is 0.1 Gyr,
and the maximum value for $A_v$ is 1.1 magnitudes. When we use only the redSED, 
the $3 \sigma$ domain is enlarged, and the maximum age and $A_v$
permitted for a burst model are 0.29 Gyr and 1.6 magnitudes respectively. 
The best fits of H3 are given by a constant star-formation model, 
0.02 $Z_{\odot}$ metallicity, with $A_v=0.35$ and age 2.4 Gyr ($\chi^2=0.85$), but
this age is in fact incompatible with the age of the universe 
(1.17(2.16) Gyr with the adopted cosmology and $q_0$=0.5(0.1)). This is not a problem 
because the 
$1 \sigma$ domain for this model encloses solutions up to $Z_{\odot}$ metallicity, 
with ages starting at 0.18 Gyr and limited by the age of the universe. 
$A_v$ ranges between 0.1 and 1.1 magnitudes at $3 \sigma$, and it increases up to 
1.6 magnitudes when using the redSED. 

H5 is well fitted by the burst and the constant star-formation models, provided that
the $A_v$ value keeps below 0.15 magnitudes typically. 
The best fit of H5 is given by a constant star-formation model of $2.5 Z_{\odot}$,
age 0.51 Gyr and $A_v=0.0$ ($\chi^2= 0.83$). The best fit by the burst model
has $Z_{\odot}$, age 0.012 Gyr and $A_v=0.0$ ($\chi^2= 0.99$).
For the burst model, the age and the reddening are well constrained.
All the solutions within $2 \sigma$ are 
compatible with very small values of $A_v$ ($A_v \ltapprox 0.3$), whatever the
model or the age of the stellar population. This result is also found when fitting
only the redSED. At $3 \sigma$, the maximum age of the burst is 0.07 Gyr,
and up to 0.14 Gyr whith the redSED. The best fit models have solar metallicities 
or higher than solar. For the constant star-formation model, the reddening is
well constrained. The ages are also well constrained (between 0.07 and 
the age of the universe), provided that we can exclude the highest metallicity. 


The best-fit SEDs were used to compute the SFR values and the absolute magnitudes 
involved in the bursts from the observed magnitudes and mean fluxes. The
lens-corrected luminosity observed within the $I_{W}$ band is 
$0.6 (1.7) 10^{45} h_{50}^{-2}$ erg s$^{-1}$ for H3, and
$2.6 (7.4) 10^{44} h_{50}^{-2}$ erg s$^{-1}$ for H5, with $q_0=0.5(0.1)$, assuming $A_v = 0$.
In the case of H3, the best fit model has $A_v \sim 0.5$, thus the corrected luminosity 
in this filter (restframe $\sim$ 1450 - 1700 \AA) is in fact 2.8 times higher. When using 
the best-fit models to scale 
the fluxes (Fig. 5), we obtain L(1500 \AA)= 1.24 (3.5) $10^{42} h_{50}^{-2}$ 
erg s$^{-1}$ \AA$^{-1}$ \
for H3 (corrected for $A_v=0.5$), with $q_0=0.5(0.1)$, and 
L(1500 \AA)= 1.9 (5.4) $10^{41} h_{50}^{-2}$ erg s$^{-1}$ \AA$^{-1}$ \ for H5 ($A_v = 0$). These
values are not strongly dependent on the model SED used for the scaling. The SFRs
derived from the best-fit star-forming models, scaled to the observed L(1500 \AA), are
$311 (110) h_{50}^{-2}$ $M_{\odot}/yr$ for H3 (corrected for $A_v=0.5$) and
$54 (19) h_{50}^{-2}$ $M_{\odot}/yr$ for H5 ($A_v = 0$), and only $\sim 1/3$ of these
values correspond to stars with $m \gtapprox 1M_{\odot}$. The main uncertainty is the
$A_v$ for H3, which could set the SFR to values $\sim 3$ times lower or higher than
the best solution quoted above. The absolute B magnitudes were derived from the observed
K magnitudes (restframe $\sim$ 3900 - 4430 \AA), using the best-fit models for a detailed scaling: 
$M_B=-22.9$ $(-24.1)$ and $M_B=-20.7$ $(-21.9)$ for H3 and H5 respectively, 
with $q_0=0.5(0.1)$. In the case of H5, this result is almost independent on the model 
SED because $A_v$ is always small; in the case of H3, the $3 \sigma$ uncertainties in 
$A_v$ translate into $\sim \pm 0.7$ magnitudes of uncertainty in $M_B$.

The shape of the continuum at wavelengths shorter than $\sim 8000 $ \AA \ (restframe $\sim 1500 $ \AA ), 
up to the I band, is relatively insensitive to the metallicity for metallicities higher than solar. 
The likelihood map is almost insensitive to age for ages below $\sim 10^6$ years, as 
expected given the sampling in stellar masses in the evolutionary tracks used by GISSEL98. 
Nevertheless, such time-scales seem irrelevant here, the best solutions being older than
this limit. These results also are generally insensitive to the choice of the IMF. Nevertheless,
the permitted regions in the likelihood maps show some dependency on the upper mass 
limit assumed for star-formation. In particular, when this limit is set to a value as
low as $10 M_{\odot}$, the permitted region for burst models is shifted towards 
younger age values for the stellar population, irrespective of the metallicity. This change on 
model details has very small influence on the SFRs derived above.

\section{Discussion and conclusions}

The two sources H3 and H5 are at the same redshift, but their stellar populations are 
noticeably different. Both of them are well fitted by burst or continuous star-formation
models. Metallicities much higher than $2.5 Z_{\odot}$ seem to be excluded by the photometric data
in the case of H3, and they have a lower probability in H5. Both galaxies are roughly
compatible with solar metallicities within the $1 \sigma$ confidence level: 
$Z \ltapprox Z_{\odot}$ for H3 and $Z \gtapprox 0.2 Z_{\odot}$ for H5. The stellar population
seen in H5 is younger than in H3, on the basis of a burst model with $A_v = 0$.
The main source of uncertainty when deriving the restframe properties of these objects 
is the value of $A_v$, and this is clearly shown in the case of H3 where the $3 \sigma$ 
uncertainties are $\sim \pm 0.5$ magnitudes. The two sources are intrinsically bright: roughly
$M^{*}_B$ for H5 and $M^{*}_B - 2 $ ($\pm 0.7$) magnitudes for H3 (here $M^{*}_B$ is the local
value, from Loveday et al. 1992). This, combined with the high
gravitational amplification and the presence of relatively strong emission lines, has 
allowed us to obtain a spectroscopic redshift for these
sources using a 4m telescope. Among the high redshift candidates in our sample behind 
A2390, 4 additional ones are at the same photometric redshift. A subsequent spectroscopic 
survey using a 8m class telescope is needed to go further on this study, especially to 
estimate more precisely the metallicities from UV absorption lines. 

When comparing our results on H3 to those by Frye \& Broadhurst (1998) and Bunker et al.
(1998), we find them in fairly good agreement, after correction for the difference in 
the amplification factor, which is 3 times higher in their lens model. The age of the 
system and the reddening value given by Bunker et al. for H3 are included within our 
$2 \sigma$ best solution region, which is a remarkable result taking into account that both the
photometric data (HST excepted) and the analysis are completely independent. The 
high amplification factor obtained by Frye \& Broadhurst (1998) corresponds to 
the maximum value attained in our model for H3, on the region neighbouring the critical line,
but the surface-averaged value in our case is 3 times lower.

Compared to other known $z \ge 3$ galaxies, H3 and H5 belong to the bright end of
the field population (Lowenthal et al. 1997; Steidel et al. 1996a and 1996b). They are 
slightly brighter than the $z \sim 4$ objects found by Trager et al. (1997) in the cluster
lens Cl0939+4713, although the amplification factor is poorly known in this case.
They are also intrinsically brighter than the $z=5.34$ galaxy found by 
Dey et al. (1998). 
These sources are not resolved in their width, where the lens inversion is limited by the resolution 
of the WFPC2 images (about $0.1''$, $0.9 (1.6) h_{50}^{-1}$ kpc with $q_0= 0.5 (0.1)$), but 
they are resolved on their length. H3 and H5 are splitted into several small, compact and
bright subclumps, all aligned towards the same direction. The total length of these emitting
regions is similar to that of the compact cores of the $2.5 \ltapprox z \ltapprox 3.5$ field sample 
by Steidel et al. (1996a). H3 is more elongated than H5, and probably more dusty. 
The linear separation between H3 and H5 ($150 (252) h_{50}^{-1} kpc$ with $q_0= 0.5 (0.1)$) 
and their peculiar morphologies strongly point towards a hierarchical merging process
as a likely scenario for the formation of the brightest spheroids. 
Their photometric SEDs are compatible with a wide range of metallicities at a
$3 \sigma$ level, all with $ Z \gtapprox 0.02 \% Z_{\odot}$. Nevertheless, as shown in
Section 4, broad band photometry does not allow a precise estimate of the metallicity.
A spectroscopic study using UV restframe absorption lines, or near-IR spectroscopy, 
is urgently needed for these purposes. A relatively high metallicity for the bright subclumps
would imply that we are actually seeing an advanced step in this merging process, where 
stars are forming from a metal enriched gas (see also Trager et al. 1997; Lowenthal et al. 1997, 
Baugh et al 1997; Moscardini et al. 1997). It is worth noting that we are dealing with 
local conditions in these star forming systems. The light detected is mainly emitted in
very small and compact regions ($\sim 1 h_{50}^{-1}$ kpc), which could be highly enriched compared
to the remainder of the source. This is fully compatible with the best fit metallicities being
$ \gtapprox 0.02 \% Z_{\odot}$ in these systems, because in this case the star formation 
becomes a local and efficient process, where the cooling rate is enhanced by metallic 
atoms allowing the formation of molecules and increasing the dust opacity. Such a process is 
discussed in details in a recent paper by Spaans \& Carollo (1998).

The uncertainty in the amplification factor for H3 and H5 is 0.3 magnitudes. This means that 
the intrinsic luminosities and SFRs are known with an accuracy of $\sim 30\%$.
This source of error has the same importance than the model uncertainties for a relatively
well constrained SED (i.e., the uncertainty of $\sim 0.5$ magnitudes in $A_v$ for H3). It is worth 
noting that H3 and H5 are multiple images, well modelled compared to other images in this or 
other cluster lenses. 
This gives an idea of the limitation arising from lens modelling when using clusters as
gravitational telescopes to access the background sources. Only the well constrained 
clusters are actually useful for this programme.

H3 and H5 are the first spectroscopically confirmed images of sources at $z \gtapprox 2$ in
this cluster, a redshift domain which is well constrained by the set of filters used here. 
The selection of high-redshift candidates using a photometric redshift
approach, including the near-IR bands, is strongly supported by the present results.
For most statistical purposes, photometric redshifts should be accurate enough to
discuss the properties of these extremely distant galaxies. Conversely, the spectroscopic
confirmation of the redshifts of such gravitationally amplified sources could help 
on the calibration and improvement of the photometric redshifts techniques up to 
much fainter limits in magnitude compared to field surveys. 

\acknowledgements
We thank Y. Mellier, G. Mathez, B. Fort, J.P. Picat and E. Hatziminaoglou 
for useful discussions about 
this particular program, and R.S. Ellis for a careful reading of the manuscript. We are 
grateful to M. Breare and all the La Palma staff for their support during the LDSS-2 runs. 
Part of this work was supported by the French {\it Centre National de
la Recherche Scientifique}, by the French {\it Programme National de 
Cosmologie} (PNC), and the TMR {\it Lensnet} ERBFMRXCT97-0172
(http://www.ast.cam.ac.uk/IoA/lensnet).
G. Bruzual thanks the OMP for its hospitality, and the MENRT and
the EEC program No. CHRX-CT92-0033 for support. We also acknowledge the DGICYT 
(Ministerio de Educaci\'on y Ciencia. Spain) (IT).

\end{document}